\documentclass[reprint,nofootinbib,superscriptaddress,amsmath,amssymb,aps,prd]{revtex4-1}
\usepackage{graphicx}
\usepackage{rotating}
\usepackage{dcolumn}
\usepackage{bm}
\usepackage{color}
\usepackage{mathptmx, textcomp}
\usepackage[latin1]{inputenc}
\usepackage{braket}
\usepackage[colorlinks,linkcolor=magenta,anchorcolor=cyan,citecolor=blue]{hyperref}
\usepackage[multiple]{footmisc}
\usepackage{booktabs}    	
\usepackage{multirow}

\newcommand{\bea}{\begin{eqnarray*}}
\newcommand{\eea}{\end{eqnarray*}}
\newcommand{\bean}{\begin{eqnarray}}
\newcommand{\eean}{\end{eqnarray}}
\newcommand{\eqs}[1]{Eqs. (\ref{#1})}
\newcommand{\eq}[1]{Eq. (\ref{#1})}
\newcommand{\meq}[1]{(\ref{#1})}
\newcommand{\grad}{\nabla}

\newcommand{\non}{\nonumber \\}

\newcommand{\hsp}{\hspace{0.1mm}}

\newcommand{\pp}{\partial}

\begin{document}

\title{The Quasinormal Modes and Isospectrality of Bardeen (Anti-) de Sitter Black Holes}

\author{Ying Zhao}
\thanks{These authors contributed equally to this work}
\affiliation{Department of Physics, Key Laboratory of Low Dimensional Quantum Structures and Quantum Control of Ministry of Education, and Synergetic Innovation Center for Quantum Effects and Applications, Institute of Interdisciplinary Studies, Hunan Normal University, Changsha 410081, Hunan, People's Republic of China}

\author{Wentao Liu}
\thanks{These authors contributed equally to this work}
\affiliation{Department of Physics, Key Laboratory of Low Dimensional Quantum Structures and Quantum Control of Ministry of Education, and Synergetic Innovation Center for Quantum Effects and Applications, Institute of Interdisciplinary Studies, Hunan Normal University, Changsha 410081, Hunan, People's Republic of China}

\author{Chao Zhang}
\affiliation{Department of Physics, Faculty of Science, Tokyo University of Science,
1-3, Kagurazaka, Shinjuku-ku, Tokyo 162-8601, Japan}

\author{Xiongjun Fang}
\email[Corresponding author: ]{fangxj@hunnu.edu.cn}
\affiliation{Department of Physics, Key Laboratory of Low Dimensional Quantum Structures and Quantum Control of Ministry of Education, and Synergetic Innovation Center for Quantum Effects and Applications, Institute of Interdisciplinary Studies, Hunan Normal University, Changsha 410081, Hunan, People's Republic of China}

\author{Jiliang Jing}
\affiliation{Department of Physics, Key Laboratory of Low Dimensional Quantum Structures and Quantum Control of Ministry of Education, and Synergetic Innovation Center for Quantum Effects and Applications, Institute of Interdisciplinary Studies, Hunan Normal University, Changsha 410081, Hunan, People's Republic of China}

\begin{abstract}
Black holes (BHs) exhibiting coordinate singularities but lacking essential singularities throughout the entire spacetime are referred to as regular black holes (RBHs). The initial formulation of RBHs was presented by Bardeen, who considered the Einstein equation coupled with a nonlinear electromagnetic field. In this study, we investigate the gravitational perturbations, including the axial and polar sectors, of the Bardeen (Anti-) de Sitter black holes. We derive the master equations with source terms for both axial and polar perturbations, and subsequently compute the quasinormal modes (QNMs) through numerical methods. For the Bardeen de Sitter black hole, we employ the 6th-order WKB approach. The numerical results reveal that the isospectrality is broken in this case. Conversely, for Bardeen Anti-de Sitter black hole, the QNM frequencies are calculated by using the HH method.
\end{abstract}
\maketitle

\section{Introduction}

Over these years, the gravitational wave searches for the coalescence of compact binaries \cite{LIGO} and the shadow images captured by the Event Horizon Telescope \cite{ETH1,ETH2} explicitly give the evidence to the existence of BHs. In general relativity, essential singularities are present in BHs, which cannot be eliminated by coordinate transformation. Modern physics tries to construct a complete theory of quantum gravity, and understand the microscopic mechanism of BHs. However, the description of the essential singularity has encountered enormous difficulties. To overcome this problem, one can consider the BHs without essential singularities are known as regular black holes. Comparing with those of singular black holes, the regular black holes have some new phenomena, and which attract much attention recently \cite{miao}. This kind of solutions was first proposed by Bardeen \cite{Bardeen}. Ay\'on-Beato and Garc\'ia found that, to obtain a regular black hole solution, the energy-momentum tensor should be the gravitational field of some magnetic monopole generated by a specific form of nonlinear electrodynamics \cite{EAAG1,EAAG2,EAAG3}. Since then, numerous researchers have contributed various singularity-free solutions \cite{regular1,regular2,regular3,regular4,regular5,regular6}. In recent years, it has attracted more and more attentions \cite{sumregular1,sumregular2,sumregular3}.

In the context of binary black holes coalescence, the process can be divided into three distinct stages: inspiral, merge and ringdown. Each stage is calculated by different methods. The inspiral can be discussed by the post-Newtonian approximation, and the merge is always calculated by numerical calculation. The final stage is equivalent to making a perturbation to an equilibrium state of the black hole. This stage corresponds to damped oscillations with complex frequencies, which always depend only on black hole properties, like mass, charge and angular momentum. These modes are called quasinormal modes (QNMs). The gravitational perturbation equations for the axial sector of Schwarzschild spacetime were initially formulated by Regge and Wheeler \cite{ReggeWheeler1957}, and later extended to the polar sector by Zerilli \cite{Zerilli1970PRL,Zerilli1970PRD}. Extensive research has been dedicated to numerically determining the quasinormal frequencies ($\omega$) for various scenarios. It is well-established that the axial and polar gravitational perturbations of asymptotically flat spacetime, such as those associated with Schwarzschild or Reissner-Nordstr\"{o}m BHs, exhibit isospectrality \cite{chandrasekharbook}. Conversely, the investigation of quasinormal modes in asymptotically Anti-de Sitter (AdS) spacetime has attracted significant interest due to the AdS/CFT duality \cite{Chan,HH}. Numerical investigations have revealed a distinct parity splitting phenomenon for Schwarzschild-AdS BHs \cite{Cardoso2001}. Furthermore, in addition to their classical properties, quasinormal modes also exhibit intriguing quantum characteristics inherent to BHs \cite{Hod1998,Maggiore2008}.

After the Bardeen solution has been presented, many works focused on the perturbation problem of the Bardeen BHs. The QNMs due to the neutral or charged scalar field perturbation for regular black holes have been discussed in Ref. \cite{Fernando2012,Flachi2013}. Ulhoa investigated the axial gravitational perturbations of a regular black hole \cite{Ulhoa2014}. To investigate the stability of nonlinear electrodynamics black holes, Moreno and Sarbach derive the perturbation equations \cite{Moreno2003}. Based on this result, Chaverra \textit{et al}. calculate the QNMs of black hole in nonlinear electrodynamics, and find that there exist a parity splitting for the alternative model to the Bardeen black hole \cite{Chaverra}. Further, Toshmato \textit{et al}. studied the electromagnetic perturbation of black hole coupled with nonlinear electrodynamics, and their results show a correspondence between the axial and polar parts of electromagnetic perturbations of electrically charged or magnetically charged black holes \cite{Toshmato1,Toshmato2}. The Dirac QNMs of regular black hole was also investigated in \cite{Li2013}.

The Bardeen BH with cosmological constant was initially introduced by Fernando \cite{Fernando2016}. In this study, we derive the master variables and corresponding master equations for the gravitational perturbations of Bardeen (Anti-) de Sitter BHs, employing the so-called A-K notation. Our analysis encompasses both the axial (odd-parity)  and polar (even-parity)  sectors. Subsequently, we numerically compute the QNMs, with necessary analysis. The structure of the paper is outlined as follows: In the subsequent section, we provide a brief overview of the Bardeen BH solutions and the construction of master equations for (axial and polar) gravitational perturbations. Section \ref{sec3} focuses on the study of QNMs for the Bardeen de Sitter BH utilizing the 6th-order WKB approach. Additionally, we present a comparative discussion regarding the quasinormal frequency splitting phenomenon. Moving on to Section \ref{sec4}, we employ the HH method to calculate the QNMs of the Bardeen Anti-de Sitter BH. Finally, Section \ref{sec5} is dedicated to summarizing our findings and providing a comprehensive discussion. Throughout our paper, we choose $c=G=1$ and ignore the factor $\kappa=8\pi$ in gravitational equations.

\section{Gravitational Perturbation of Bardeen (Anti-) de Sitter Black Hole}\label{sec2}

First we briefly introduce the Bardeen de Sitter BH following the work in Ref. \cite{Fernando2016}, the action is given by
\begin{equation}
S=\int d^4x \sqrt{-g}\left[\frac{\left(R-2\Lambda\right)}{16\pi }-\frac{1}{4\pi}\mathcal{L}(F) \right],
\end{equation}
where $\mathcal{L}(F)=\frac{3}{2 \alpha q^2}\left(\frac{\sqrt{2q^{2}F}}{1+\sqrt{2q^2F}}\right)^{5/2}$ is the Lagrangian of the nonlinear electromagnetic field strength $F$, and $R$, $\Lambda$ are the scalar curvature, the cosmological constant, respectively. The parameter $\alpha$ in  $\mathcal{L}(F)$ is related to the magnetic charge $q$ and the mass $M$  of the space time as follows: $\alpha=q/2M$.  The field strength $F$ can be written as
\begin{equation}
F=\frac{1}{4}F_{\mu\nu}F^{\mu\nu}
\end{equation}
where $F_{\mu\nu}=\nabla_{\mu}A_{\nu}-\nabla_{\nu}A_{\mu}$. The only non-vanishing component of $F_{\mu\nu}$ for spherically symmetric spacetime is $F_{23}=q\sin\theta$, and then $F=q^2/2r^4$.

Taking the variation of the action, the equation of motion can be expressed as
\begin{eqnarray}\label{motion}
& R_{\mu\nu}-\frac{1}{2}Rg_{\mu\nu}+\Lambda g_{\mu\nu}
=2\left(\frac{\partial \mathcal{L}(F) }{\partial F}F_{\mu\lambda }F_{\nu}\hsp^{\lambda}-g_{\mu\nu}\mathcal{L}(F)\right) \non
& \nabla_{\mu}\left(\frac{\partial \mathcal{L}(F) }{\partial F}F^{\nu\mu}\right)=0.
\end{eqnarray}
The static spherically symmetric solution is given by \cite{Fernando2016}
\begin{equation}
ds^2=-f(r)dt^{2}+f(r)^{-1}dr^{2}+r^{2}\left(d\theta ^{2}+\sin^{2}\theta d\varphi^{2}\right)
\end{equation}
where $f(r)=1-\frac{2Mr^2}{(r^2+q^2)^{\frac{3}{2}}}-\frac{\Lambda r^{2}}{3}$, here $M$, $q$ are the total mass and the magnetic charge of the Bardeen de Sitter BH. The $\Lambda>0$ case represents the Bardeen de Sitter BH, and the $\Lambda<0$ case represents the Bardeen Anti-de Sitter BH.

The decomposition of perturbed metric was first presented by Regge and Wheeler. Considering that the perturbed metric can be written as
\begin{eqnarray}
g_{ab}=\bar g_{ab}+h_{ab}
\end{eqnarray}
where $\bar g_{ab}$ denotes the metric of the background spacetime. For spherically symmetric spacetime, it is well-known that $h_{ab}$ can be decomposed to the axial part and polar part under the Regge-Wheeler gauge. In this paper, we use A-K notation to decompose $h_{ab}$ \cite{Thompson2016}, which helps to give the perturbed source terms. Note that when calculating the QNMs, we do not need the source term. However, here we present the master equation with source term, which can provide some basis to study the self-force or extreme mass ratio inspiral problems for Bardeen (Anti-)de Sitter black holes. Considering that the complete basis on the two-sphere are constructed by 1-scalar spherical harmonic, $Y_{lm}=Y_{lm}(\theta,\varphi)$, three pure-spin vector harmonics and six tensor harmonics \cite{Thorne1980}. The vector harmonics are defined as
\begin{eqnarray}
Y_a^{E,lm}=r\grad_a Y^{lm},~~~~ Y_a^{B,lm}=r\epsilon_{ab}\hsp^c n^b\grad_c Y^{lm},~~~~ Y_a^{R,lm}=n_aY^{lm},\non
\end{eqnarray}
and the tensor harmonics are defined as
\begin{eqnarray}
T_{ab}^{T0,lm}&=&\Omega_{ab}Y^{lm},~~~~T_{ab}^{L0,lm}=n_an_b Y^{lm}, \non
T_{ab}^{E1,lm}&=&r~n_{(a}\grad_{b)}Y^{lm},~~~~T_{ab}^{B1,lm}=r~n_{(a}\epsilon_{b)c}\hsp^d n^c\grad_d Y^{lm}, \non
T_{ab}^{B2,lm}&=&r^2\Omega_{(a}\hsp^c\epsilon_{b)e}\hsp^d n^e\grad_c\grad_d Y^{lm},\non
T_{ab}^{E2,lm}&=&r^2\left(\Omega_a\hsp^c\Omega_b\hsp^d-\frac{1}{2}\Omega_{ab}\Omega^{cd}\right)\grad_c\grad_dY^{lm}.
\end{eqnarray}
where $v_a=(-1,0,0,0)$ and $n_a=(0,1,0,0)$ are two orthogonal co-vectors, $\Omega_{ab}$ is the projection operator defined as $\Omega_{ab}=r^2 \text{diag}(0,0,1,\sin^2\theta)$, $\epsilon_{abc}$ represent the spatial Levi-Civita tensor $\epsilon_{abc}\equiv v^d\epsilon_{dabc}$, and now $\epsilon_{tr\theta\varphi}=r^2\sin\theta$. Now, $h_{ab}$ can be decomposed as
\begin{align}
\label{hab}
h_{ab}^{lm}&= \mathrm{A~}v_av_bY^{lm}+2\mathrm{B~}v_{(a}Y_{b)}^{E,lm}+2\mathrm{C~}v_{(a}Y_{b)}^{B,lm}+2\mathrm{D~}v_{(a}Y_{b)}^{R,lm}\non
&~~~~+\mathrm{E~}T^{T0,lm}_{ab}
 +\mathrm{F~}T_{ab}^{E2,lm}+\mathrm{G~}T^{B2,lm}_{ab}+2\mathrm{H~}T^{E1,lm}_{ab}\non
&~~~~+2\mathrm{J~}T^{B1,lm}_{ab}+\mathrm{K~}T^{L0,lm}_{ab} ,
\end{align}
where coefficients of each term in the above equation are all scalar functions of $t$ and $r$. Hence, the polar part and the axial part of $h_{ab}$ can be written as
\begin{eqnarray}
h_{ab}^{polar}=\left(\begin{array}{cccc}
\mathrm{A}Y^{lm}&-\mathrm{D}Y^{lm}&-r \mathrm{B}\partial_{\theta} Y^{lm}& -r \mathrm{B}\partial_{\phi} Y^{lm} \\
Sym&\mathrm{K}Y^{lm}&-r  \mathrm{H}\partial_{\theta} Y^{lm} & r  \mathrm{H}\partial_{\phi} Y^{lm} \\
 Sym & Sym & r^{2}~\mathrm{Z}^{lm} & r^2\mathrm{F}X^{lm} \\
Sym & Sym & Sym & r^{2} \sin^2\theta~\mathrm{Z}^{lm}
\end{array}\right)
\end{eqnarray}
\begin{align}
h_{ab}^{axial}=\left(\begin{array}{cccc}
0&0&r~csc\theta\mathrm{C}\partial_{\phi} Y^{lm}& -r ~sin\theta\mathrm{C}\partial_{\theta
} Y^{lm} \\
0&0&-r~csc\theta\mathrm{J}\partial_{\phi} Y^{lm} & r ~ \sin\theta\mathrm{J}\partial_{\theta} Y^{lm} \\
Sym & Sym & -r^{2}csc\theta\mathrm{G}X^{lm} & r^2\sin\theta\mathrm{G}W^{lm} \\
Sym & Sym & Sym & r^{2} \sin\theta\mathrm{G}X^{lm}
\end{array}\right)
\end{align}

where
\begin{eqnarray}
\mathrm{W}^{lm}&=&\left[\partial_{\theta}^{2}+\frac{1}{2}l\left(l+1\right)\right]Y^{lm}\non
\mathrm{X}^{lm}&=&\left[\partial_{\theta}\partial_{\phi}-cot\theta\partial_{\phi}\right]Y^{lm}\non
\mathrm{Z}^{lm}&=&\left(\mathrm{E}Y^{lm}+\mathrm{F}W^{lm}\right)
\end{eqnarray}

Assuming that the value of charge $q$ is fixed, taking the variation of \eq{motion} we obtain
\begin{align}
\label{Eab}
E_{ab} &= \Box h_{ab}+\nabla_a\nabla_bh^{c}_{~c}-2\nabla_{(a}\nabla^ch_{b)c}
+2R^{~c~d}_{a~b}h_{cd}\non
&~~~~-(R_a\hsp^c h_{bc}+R_b\hsp^c h_{ac})
+g_{ab}(\nabla^c\nabla^dh_{cd}-\Box h^{d}_{~d})\non
&~~~~-g_{ab}R^{cd}h_{cd}+R h_{ab}-2\Lambda h_{ab}+2\delta T^\text{Bardeen}_{ab}
\end{align}
where
\begin{align}
\delta T^\text{Bardeen}_{ab}=2\delta\left(\frac{\partial \mathcal{L}(F) }{\partial F}F_{ac }F_{b}\hsp^{c}-g_{ab}\mathcal{L}(F)\right)
\end{align}

Project the above equations to the A-K directions, one can get the projection equations $E_\mathrm{A}-E_\mathrm{K}$. The explicit expression of $E_\mathrm{A}-E_\mathrm{K}$ will be placed in Appendix \ref{appendixA}.

For spherically symmetric spacetimes, there are several gauge choices \cite{ReggeWheeler1957}. In this paper, we choose the RW gauge, i.e., choose the gauge vector $\xi_a$ as
\begin{align}
\xi_a&=\left(r\mathrm{B}+\frac{r^{2}}{2}\frac{\partial}{\partial t}\mathrm{F}\right)v_{a}Y^{lm}
+\left(r\mathrm{H}-\frac{r}{2}\frac{\partial}{\partial r}\mathrm{F}\right)Y^{R,lm}_{a}\non
&~~~~+\frac{r}{2}\mathrm{F}Y^{E,lm}_{a}+\frac{r}{2}\mathrm{G}Y^{B,lm}_{a}
\end{align}
which yields $\mathrm{B}=\mathrm{F}=\mathrm{H}=\mathrm{G}=0$. Then, the polar and the axial sectors of $h_{ab}$ become
\begin{eqnarray}
h_{ab}^{polar}=\left(\begin{array}{cccc}
\mathrm{A}Y^{lm}&-\mathrm{D}Y^{lm}&0& 0 \\
-\mathrm{D}Y^{lm}&\mathrm{K}Y^{lm}&0&   0 \\
0& 0& r^{2}~\mathrm{E}Y^{lm} & 0 \\
0 & 0 & 0 & r^{2} \sin^2\theta~\mathrm{E}Y^{lm}
\end{array}\right) \non
\end{eqnarray}
\begin{align}
h_{ab}^{axial}=\left(\begin{array}{cccc}
0&0&r~ csc\theta\mathrm{C}\partial_{\phi} Y^{lm}& -r ~sin\theta\mathrm{C}\partial_{\theta
} Y^{lm} \\
0&0&-r~ csc\theta\mathrm{J}\partial_{\phi} Y^{lm} & r~  \sin\theta\mathrm{J}\partial_{\theta} Y^{lm} \\
Sym & Sym & 0 & 0 \\
Sym & Sym & 0 & 0
\end{array}\right) \non
\end{align}

Then, we construct the master variables and master equations for axial part or polar part respectively. For axial part, let
\begin{eqnarray}
\psi^{(-)}=f \mathrm{J}
\end{eqnarray}
and $\psi^{(-)}$ satisfied
\begin{eqnarray}
\label{axialeq}
\left(f^2\frac{\pp^2}{\pp r^2}-\frac{\pp^2}{\pp t^2}+f'f\frac{\pp }{\pp r}
-V^{(-)}\right)\psi^{(-)}=S^{(-)}
\end{eqnarray}
with
\begin{eqnarray}
V^{(-)}=\frac{f}{r^2}\left(-2+\lambda +2r^{2}\Lambda+2f+rf'+r^2f''+4r^2\kappa\mathcal{L}  \right). \non
\end{eqnarray}
and
\begin{eqnarray}
S^{(-)}=f^2E_\mathrm{J}-\frac{1}{2}rf\left(f\frac{\pp}{\pp r}E_\mathrm{G}+f'E_\mathrm{G}\right).
\end{eqnarray}
where $\lambda=l(l+1)$.

For polar part, we choose the gauge invariants $\chi$ and $\varphi$ are defined as \cite{Gaugein}
\begin{eqnarray}
\chi&=& -\frac{1}{2}e^{2\Lambda(r)}\mathrm{E}, \non
\varphi&=& \frac{1}{2}\mathrm{K}-\frac{1}{2}(r\Lambda'(r)+1)
e^{2\Lambda(r)}\mathrm{E}-\frac{r}{2}e^{2\Lambda(r)}\frac{\partial}{\partial r}\mathrm{E}.
\end{eqnarray}
Note that here $\Lambda(r)$ is the metric function mentioned in \cite{Gaugein}, rather than the cosmological contant. Then the master variable $\psi^{(+)}$ can be constructed as
\begin{eqnarray}
\psi^{(+)}=\frac{\chi}{f}-\frac{2\varphi}{-2+\lambda+2r^{2}\Lambda+2f+2rf'+4r^2\kappa\mathcal{L}},
\end{eqnarray}
which satisfied the polar sector equation as
\begin{eqnarray}
\label{polareq}
\left(f^2 \frac{\pp^2}{\pp r^2}-\frac{rN}{\sigma \tau}\frac{\pp^2}{\pp t^2}+c_1\frac{\pp}{\pp r}+c_0\right)\psi^{(+)}=S^{(+)}
\end{eqnarray}
with
\begin{align}
\label{polarsource}
S^{(+)}&=\frac{r}{2f\sigma}\frac{\pp }{\pp r}E_\mathrm{A}
-\frac{M_F}{2\sigma}\frac{\pp}{\pp r}E_\mathrm{F}-N_AE_\mathrm{A}+N_FE_\mathrm{F}\non
&~~~~+\frac{rN}{2\sigma\tau}\left(\frac{r}{f\sigma}\frac{\pp}{\pp t}E_\mathrm{D}+E_\mathrm{H}+\frac{1}{2}E_\mathrm{K}\right).
\end{align}
where the parameters $N$, $\sigma$, $\tau$, $c_1$, $c_0$ and $M_F$, $N_A$, $N_F$ are all the functions of background. The explicit expression of these functions can be found in Appendix \ref{appendixB}. It is easy to check that the above results can reduce to that of GR when $q=0$.

Calculating the QNM, we set the source term vanish, and consider the variable $\psi^{(\pm)}=\psi^{(\pm)}(r)e^{-i\omega t}$. {The axial sector \eq{axialeq} can be directly written as the Schr\"{o}dinger-like equation with an effective potential,
\begin{eqnarray}
\label{axialeq2}
\frac{d^{2}{\psi^{(-)}(r)}}{dr^{2}_{\ast}}+\left[\omega^{2}-V^{(-)}\right]\psi^{(-)}(r)=0,
\end{eqnarray}
with
\begin{eqnarray}
\frac{dr}{dr^*}=f(r)
\end{eqnarray}
where $\omega$ is the quasinormal frequencies, which represent the black hole will oscillate under the gravitational perturbations. However, for polar sector, denoting \eq{polareq} as
\begin{eqnarray}
\label{eta123}
\left[\eta_1\frac{\partial^2}{\partial r^2}+\eta_2 \frac{\partial}{\partial r}+\left(-\frac{\partial^2}{\partial t^2}+\eta_3\right)\right]\psi^{(+)}(r)=0
\end{eqnarray}
where
\begin{eqnarray}
\eta_1=f^2\frac{\sigma\tau}{r N},~~~~\eta_2=c_1\frac{\sigma\tau}{r N},~~~~\eta_3=c_0\frac{\sigma\tau}{r N}
\end{eqnarray}
and following the standard steps presented in \cite{Chao2022}, the standard Schr\"{o}dinger-like equation is given by
\begin{eqnarray}
\label{polareq2}
\frac{d^{2}{Z}}{dz^{2}}+\left[\omega^{2}-V^{(+)}\right]Z=0,
\end{eqnarray}
where
\begin{eqnarray}
Z=\eta_1^{-\frac{1}{4}}exp\left(\frac{1}{2}\int\frac{\eta_2}{\eta_1}dr\right)\psi^{(+)}(r)
\end{eqnarray}
and
\begin{eqnarray}
\label{}
\frac{dr}{dz}=\eta_1^{\frac{1}{2}}
\end{eqnarray}
meanwhile the potential in \eq{polareq2} reads as
\begin{eqnarray}
V^{(+)}=& -\frac{\sqrt{\eta_1}}{2}\frac{\partial}{\partial r}\left(\frac{d\sqrt{\eta_1}}{dr}-\frac{\eta_2}{\sqrt{\eta_1}}\right)' \non
& +\frac{1}{4}\left(\frac{d\sqrt{\eta_1}}{dr}-\frac{\eta_2}{\sqrt{\eta_1}}\right)^2-\eta_3.
\end{eqnarray}

\section{The Quasinormal Modes for Bardeen de Sitter Black Hole}\label{sec3}
\subsection{6th-order WKB Approach}

In this section, we calculate the QNMs of the gravitational perturbation of the Bardeen de Sitter spacetime as a function of the magnetic charge $q$ and the cosmological constant by applying the 6th-order WKB approach. The WKB approximation is one of the most widely used method for calculating the QNMs of BHs. This method is designed to solve the problem of scattered waves near the peak of the potential barrier $V(r_0)$ in quantum mechanics which employ a semianalytic technique \cite{BFSCMW, SICMW, SI1987, konoplya2003, JMatyjasek, Guo:2021bcw}. Since the complete expression of the potential $V^{(+)}$ in \eq{polareq2} is too complex, we expand the wave equations up to the fourth order of $q$ in the following calculations. Hence, $f(r)$ and $\mathcal{L}$ are considered to be rewritten as
\begin{align}
\label{frq4}
f(r)=& 1-\frac{2M}{r}-\frac{\Lambda r^{2}}{3}+\frac{3Mq^{2}}{r^{3}}-\frac{15Mq^{4}}{4r^{5}}\\
\label{Lq4}
\mathcal{L}=& \frac{3Mq^{2}}{r^{5}}-\frac{15Mq^{4}}{2r^{7}}
\end{align}
In next subsection, the numerical result show that the 4th order expansion of $q$ has enough accuracy for our discussion. However, in order to improve the computational accuracy, we expand $f(r)$ and $\mathcal{L}$ to the 10th order of $q$ in our calculation program.

Here we use the 6th-order WKB approach to calculate the QNMs for Bardeen de Sitter BH. For general potential $V(r)$, the sixth order formula is given by
\begin{equation}
\label{WKBexpres}
i\frac{\omega^{2}-V_0}{\sqrt{-2V_0''}}-\Lambda_{2}-\Lambda_{3}-\Lambda_{4}-\Lambda_{5}-\Lambda_{6}=n+\frac{1}{2}
\end{equation}
where
\begin{equation}
V_0=V|_{r=r_{max}},\quad V_0''=\frac{d^2V}{dr_{\ast}^2}|_{r=r_{max}},
\end{equation}
$r_{max}$ corresponding to the maximum value of the potential $V$, $n$ is the overtone number which can be set as $n=0,1,2...$. The explicit expressions of $\Lambda_{2}$ and $\Lambda_{3}$ can be found in \cite{SICMW,SI1987}, and $\Lambda_{4}$, $\Lambda_{5}$ and $\Lambda_{6}$ can be found in \cite{konoplya2003}.

\subsection{Numerical Results}

The gravitational modes exist for $l\geq2$. In de Sitter spacetime, we set $M=1$. Fig. \ref{fig1} describes the QNMs of axial gravitational perturbation when expanding to various order of $q$. For polar gravitational perturbation, the results also indicate similar convergent behavior. It shows that expanding to the 4th order of $q$ is sufficient, i.e. \eqs{frq4} and \meq{Lq4} are valid in our calculation.

\begin{figure}[h]
	\centering
	\includegraphics[width=1.0\linewidth]{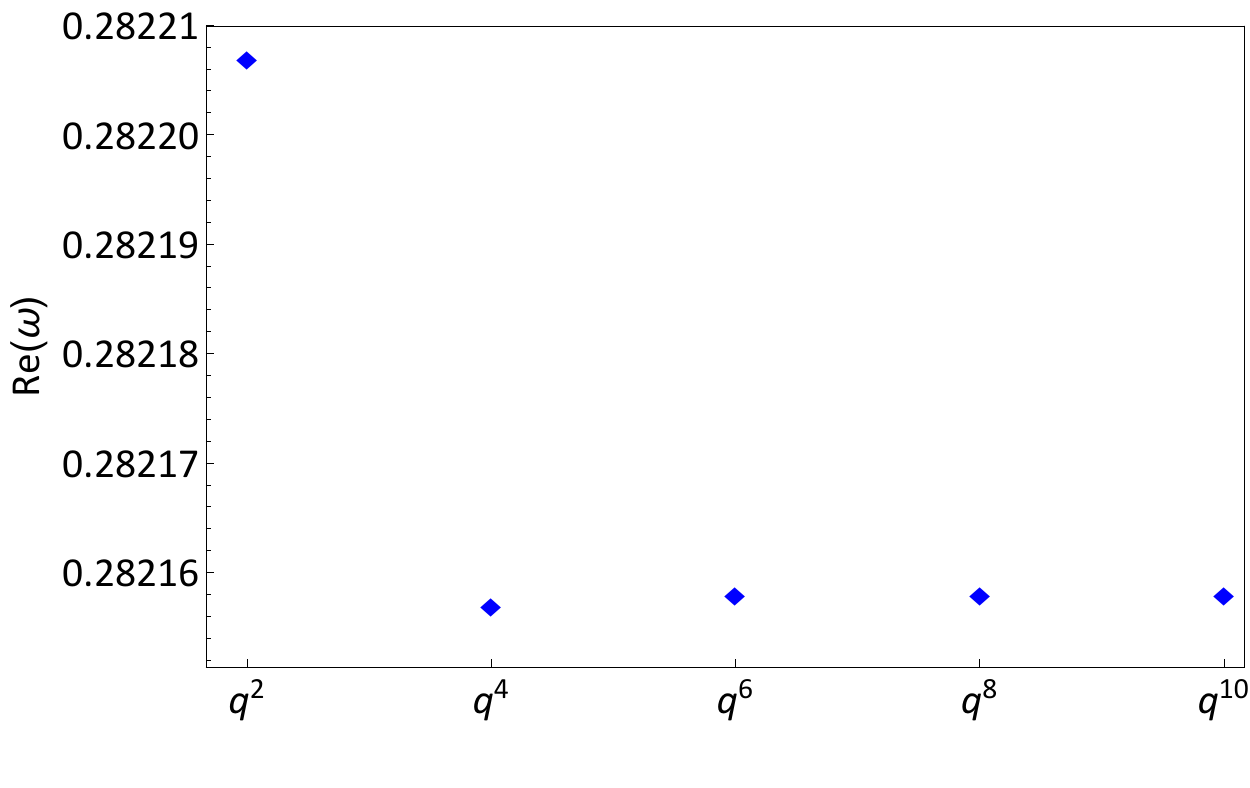}
    \includegraphics[width=1.0\linewidth]{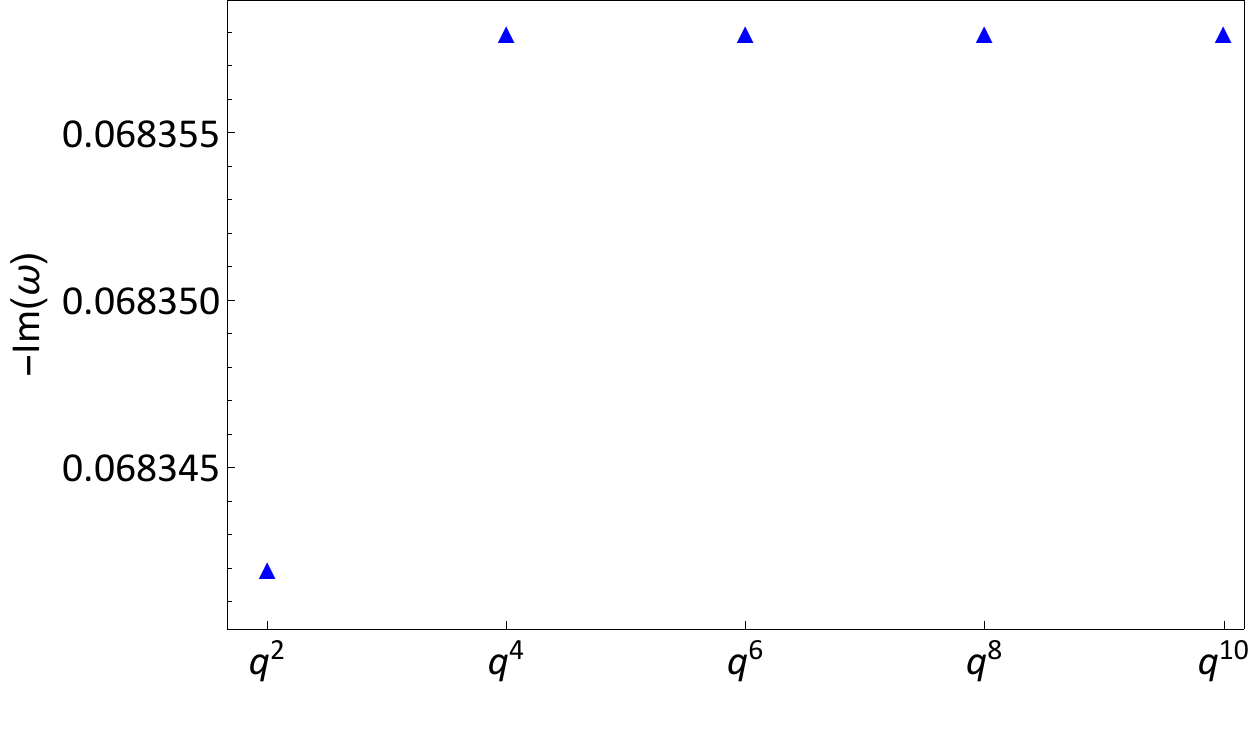}
	\caption{QNMs of axial gravitational perturbation for varying the expanding order of $q$, here we set $l=2$, $n=0$, $q=0.2$ and $\Lambda=0.05$.}
	\label{fig1}
\end{figure}

Fig. \ref{fig2} and Fig. \ref{fig3} show the QNMs of the axial and the polar parts of the gravitational perturbation for $l=2$ and $n=0$, respectively. The figures reveal that, with the increase of $q$, the real part will increase but the magnitude of imaginary part will decrease first and then increase. The specific data of QNM frequencies are place in Appendix \ref{appendixC}. The data in Table \ref{tab:2} indicate that there exist breaking of isospectrality of QNMs in Bardeen de Sitter BH.

Note that one should confirm this breaking of isospectrality is not caused by the WKB approach method. We calculate the QNMs by using various order of WKB approach, and find that the error caused by considering various order of WKB approach is much smaller than the breaking of isospectral when $q=0.6$. Therefore, we consider that the axial and the polar gravitational perturbations of Bardeen de Sitter BHs are not isospectral.

\begin{figure}[h]
	\centering
	\includegraphics[width=1.0\linewidth]{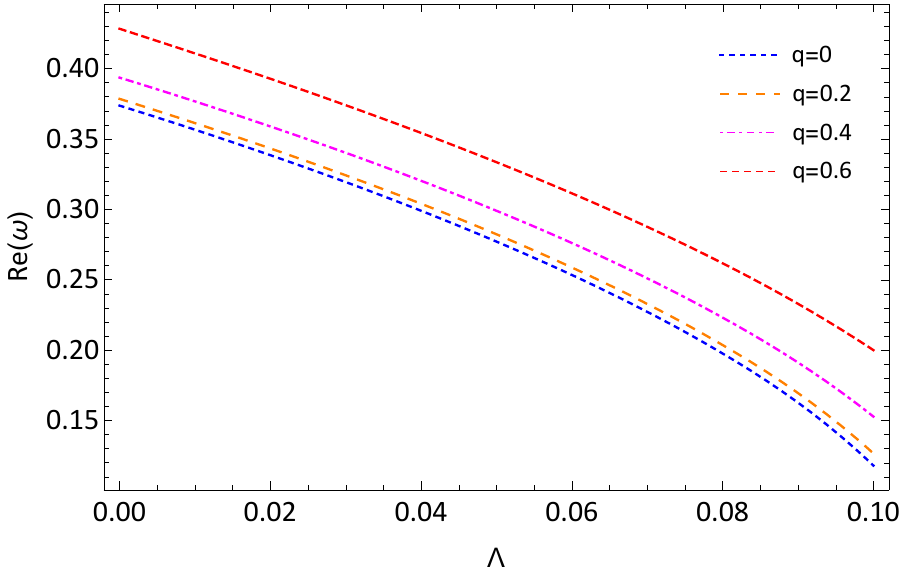}
	\includegraphics[width=1.0\linewidth]{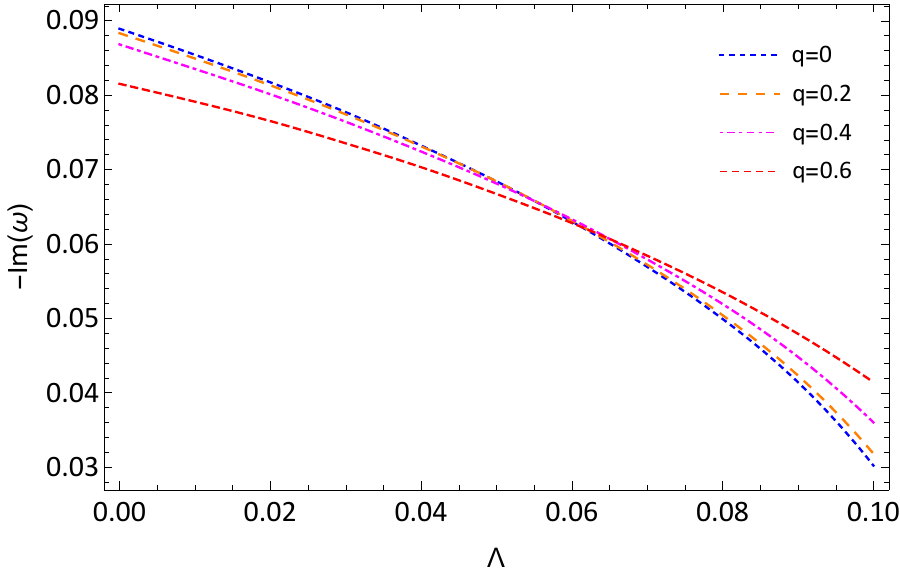}
	\caption{The QNM frequencies of the axial gravitational perturbation of the Bardeen de Sitter BHs for $l=2,n=0$.}
	\label{fig2}
\end{figure}

\begin{figure}[h]
	\centering
	\includegraphics[width=1.0\linewidth]{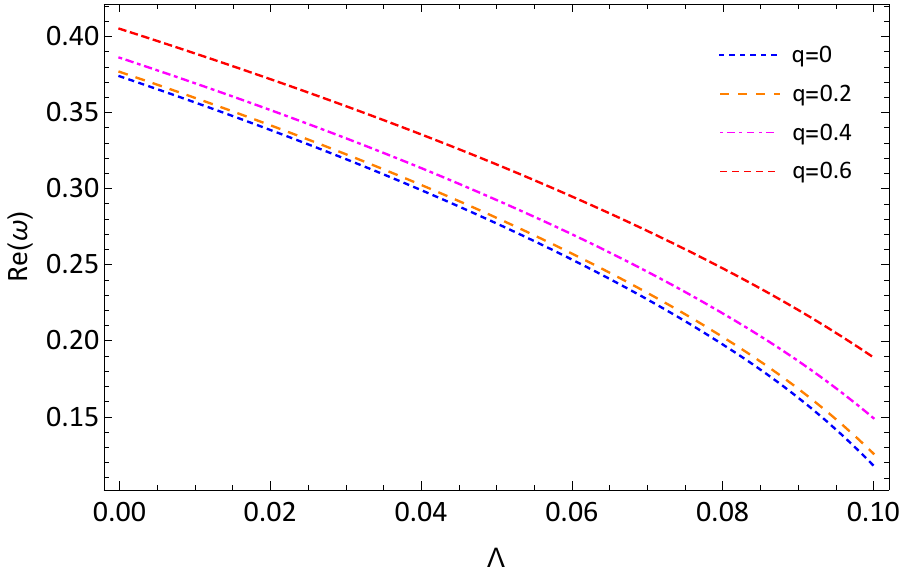}
	\includegraphics[width=1.0\linewidth]{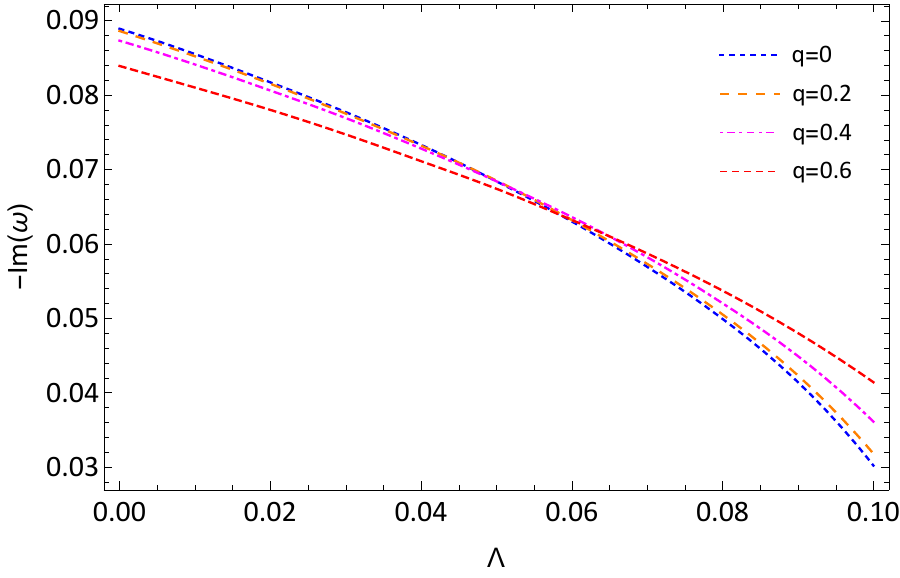}
	\caption{The QNM frequencies of the polar gravitational perturbation of the Bardeen de Sitter BHs for $l=2,n=0$.}
	\label{fig3}
\end{figure}

In Fig. \ref{fig4}, we show the relative gap between the axial gravitational and the polar gravitational QNM frequencies, where $\Delta\omega$ is defined as
\begin{equation}
\Delta\omega=\frac{\omega^{\text{odd}}-\omega^{\text{even}}}{\omega^{\text{odd}}}\times  100{\%}.
\end{equation}
This result clearly shows that with the increase of the charge $q$, isospectrality will be broken in Bardeen de Sitter spacetime. However, when $l$ is sufficiently large, our numerical results show that the frequency spectra for axial and polar gravitational perturbations converge, which implies that the isospectrality will tend to be preserved for large $l$'s.

\begin{figure}[h]
	\centering
	\includegraphics[width=1.0\linewidth]{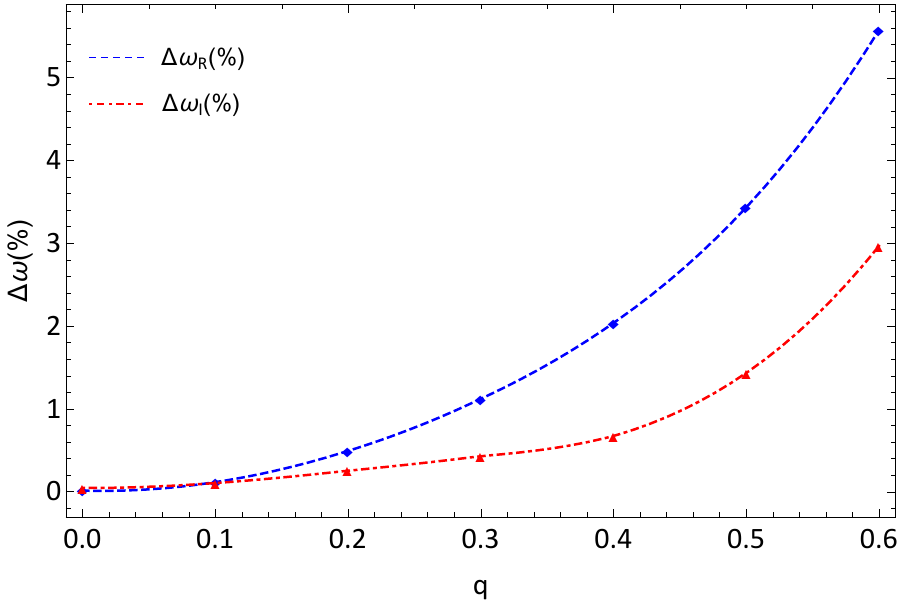}
	\caption{The relative gap between the axial and the polar QNMs of Bardeen de Sitter BHs with $\Lambda=0.02$ and $l=2, n=0$.}
	\label{fig4}
\end{figure}

\section{The Quasinormal Modes for Bardeen Anti-de Sitter Black Hole}\label{sec4}
\subsection{HH method}

For Bardeen  Anti-de Sitter BH, we use the HH method to calculate the QNMs \cite{HH,Cardoso2001}. A brief introduction of this method is as follows. In the HH method, the condition $M=1$ does not hold any more.  For axial sector, we write $\phi^{(-)}$ for a generic wavefunction as
\begin{equation}
\phi^{(-)}=e^{i\omega r_*}\psi^{(-)},
\end{equation}
the Schr{\"o}dinger-like equation becomes as
\begin{equation}
\label{axialtran}
f(r)\frac{\partial^{2}{\phi^{(-)}}}{\partial{r}^{2}}+\left[f'(r)-2i\omega\right]\frac{\partial \phi^{(-)}}{\partial r}-\frac{V}{f(r)}\phi^{(-)}=0.
\end{equation}
And for the polar sector, from \eq{polareq2} we consider that $\phi^{(+)}$ can be written as
\begin{equation}
\phi^{(+)}=e^{i\omega z}Z
\end{equation}
and satisfied
\begin{equation}
\label{polartran}
g(r)\frac{\partial^{2}{\phi^{(+)}}}{\partial{r}^{2}}+\left[g'(r)-2i\omega\right]\frac{\partial \phi^{(+)}}{\partial r}-\frac{V}{g(r)}\phi^{(+)}=0
\end{equation}
where
\begin{equation}
\frac{\partial r}{\partial z}=f(r)\sqrt{\frac{\sigma\tau}{r N}}=g(r)
\end{equation}

Then introducing a transformation $x=1/r$ to restrict the studied region $r_{h}<r<\infty$ to a finite region $0<x<x_{h}$, expanding \eqs{axialtran} and \meq{polartran} at the horizon $x_h$, we obtain
\begin{equation}
\label{equationhh}
s(x)\frac{d^{2}}{dx^{2}}\phi(x)^{(\pm)}+\frac{t(x)}{x-x_{h}}\frac{d}{dx}\phi(x)^{(\pm)}+\frac{u(x)}{\left(x-x_{h}\right)^{2}}\phi(x)^{(\pm)}=0.	
\end{equation}
The coefficient functions $s(x)$, $t(x)$ and $u(x)$ can be expanded at horizon $x=x_h$,
\begin{equation}
s(x)=\sum_{n=0}^{\infty}s_n(x-x_h)^n
\end{equation}
and similarly for $t(x)$ and $u(x)$. Note that $u(x_h)=0$ yields that $u_0=0$. After that, we consider the solution for \eqs{axialtran} and \meq{polartran},
\begin{equation}
\phi(x)=\sum^{\infty}_{n=0}a_{n}(\omega)\left(x-x_{h}\right)^{n},
\end{equation}
and \eq{equationhh} gives
\begin{equation}
a_{n}(\omega)=-\frac{1}{P_{n}}\sum_{k=0}^{n-1}\left[k(k-1)s_{n-k}+k t_{n-k}+u_{n-k}\right]a_{k},
\end{equation}
where
\begin{equation}
P_{n}=n(n-1)s_{0}+n t_{0}.
\end{equation}
Using the boundary condition $\phi=0$ at infinity, i.e., $x=0$,
\begin{equation}
\label{hhomega}
\sum^{\infty}_{n=0}a_{n}(\omega)\left(-x_{h}\right)^{n}=0
\end{equation}
Now the problem is reduced to that of finding a numerical solution of \eq{hhomega}. Since one can not get a full sum from $0$ to infinity, the summation should be cut off at an appropriate position $N$. Taking a partial sum from $0$ to $N$, the numerical roots for $\omega_N$ of \eq{hhomega} can be evaluated by some numerical method. Then we move onto the case that taking a partial sum from $0$ to $N+1$, and similarly determine $\omega_{N+1}$. Comparing $\omega_N$ and $\omega_{N+1}$ can help us determine $N$ to achieve certain computational accuracy. It shows that when $N=50$ for axial perturbation or $N=150$ for polar perturbation, the calculation accuracy have 3 significant digits. Note that in Bardeen Anti-de Sitter spacetime, we set $\Lambda=-3$.

\subsection{Numerical Results of QNMs}

\begin{figure}[h]
	\centering
	\includegraphics[width=1.0\linewidth]{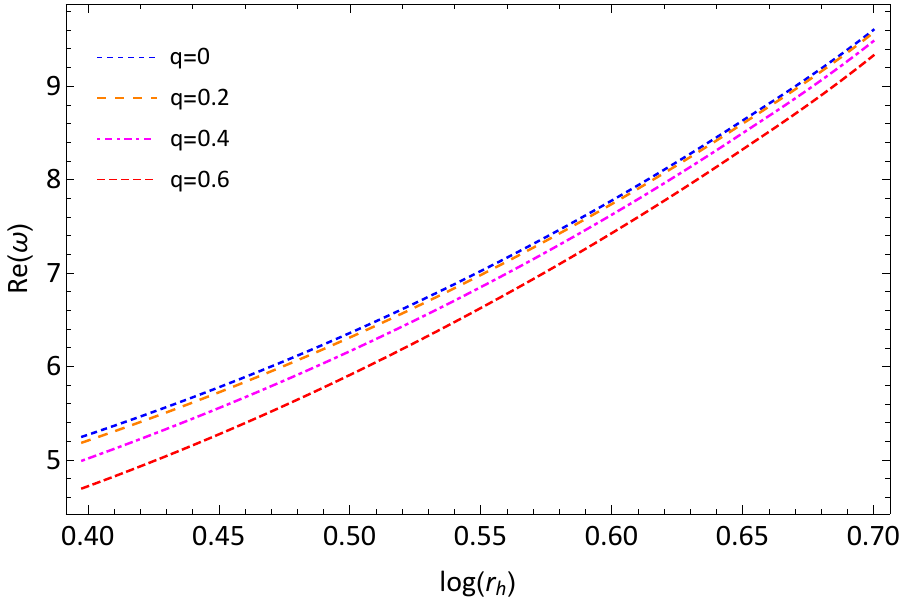}
	\includegraphics[width=1.0\linewidth]{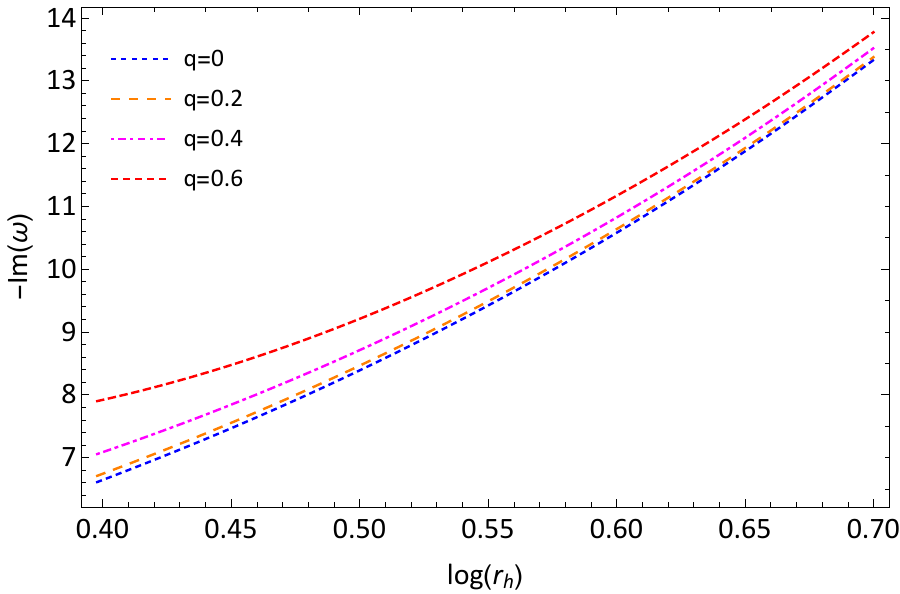}	
	\caption{The QNM frequencies of the axial gravitational perturbation of the Bardeen Anti-de Sitter BHs with $l=2$ and $n=1$.}
	\label{fig5}
\end{figure}
\begin{figure}[h]
	\centering
	\includegraphics[width=1.02\linewidth]{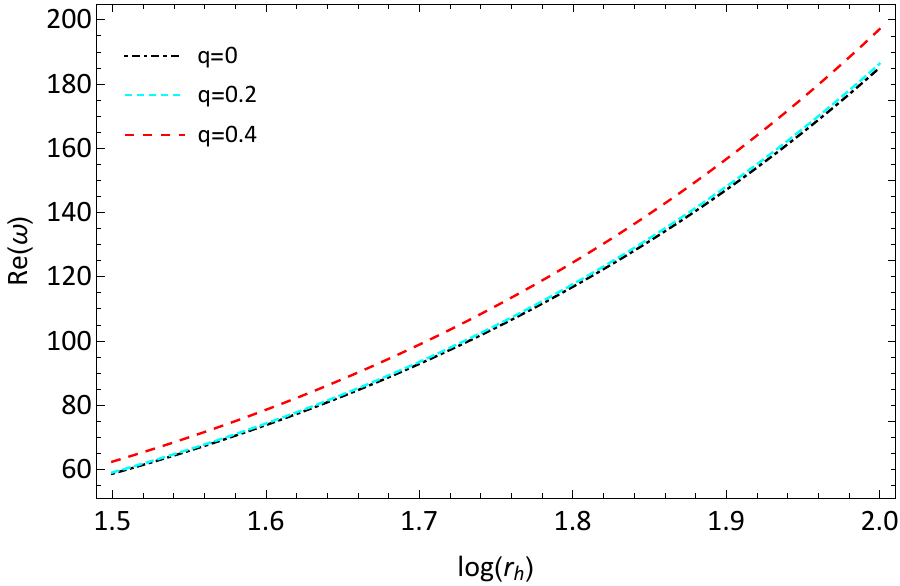}
	\includegraphics[width=1.02\linewidth]{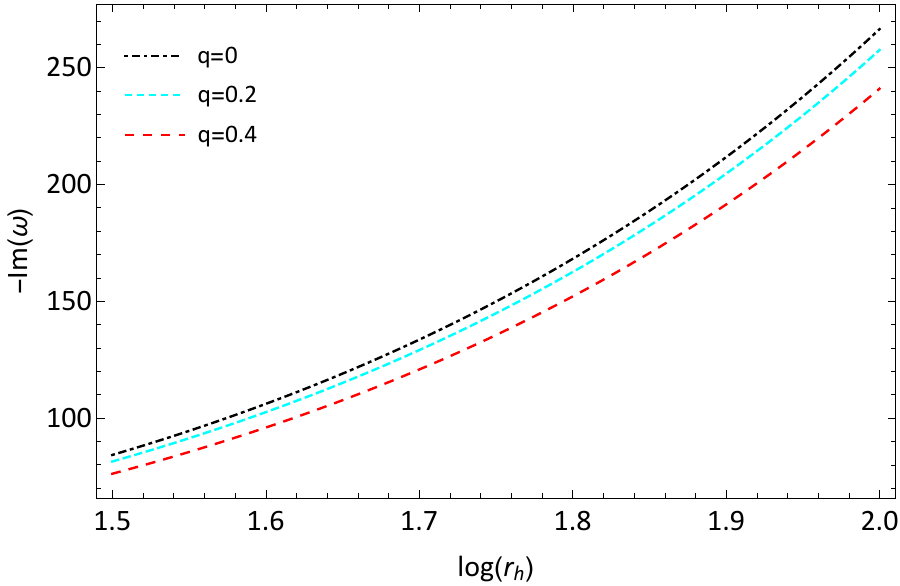}	
	\caption{The QNM frequencies of the polar gravitational perturbation of the Bardeen Anti-de Sitter BHs with $l=2,n=0$.}
	\label{fig6}
\end{figure}

Here we use HH method to calculate the QNMs of the Bardeen Anti-de Sitter BH, and study the influence of magnetic charge $q$ on QNMs. The range of $r_h$ is limited to $r_{h}\in\left[1,100\right]$. The specific data are placed in Appendix \ref{appendixC}. Fig. \ref{fig5} and Fig. \ref{fig6} show the axial and the polar parts of the gravitational QNM frequencies for Bardeen Anti-de Sitter spacetime. For the axial gravitational perturbation, we only consider the range $\log{r_h}\in [0.4, 0.7]$, to obviously show that as the charge $q$ increase, the real part of $\omega$ will decrease, and the magnitude of imaginary part of $\omega$ will increase}. While for the polar gravitational perturbation, we only consider the range $\log{r_h}\in [1.5, 2.0]$ where the deviations which due to the presence of charge $q$ can be clearly seen.

\section{Conclusion And Discussion}\label{sec5}
In this study, we investigate the gravitational perturbations of the Bardeen BH in the presence of a cosmological constant. Both the axial (odd-parity) and polar (even-parity) sectors are considered, we provide the detailed construction process for the decoupled equations with source terms. Then we derive Schr{\"{o}}dinger-type equations with effective potentials given by equations \eqref{axialeq2} and \eqref{polareq2}. Subsequently, we apply the WKB approach to analyze the quasinormal modes (QNMs) for the Bardeen de Sitter spacetime. The obtained QNM frequencies are presented in Tables \ref{tab:1} and \ref{tab:2}. We explore various combinations of values for $q$ and $\Lambda$ while fixing $l$ and $n$, and vice versa.
The results reveal deviations between the QNM frequencies in the polar sector and those in the axial sector. To ensure that these deviations are not due to numerical errors, we conduct additional calculations using different orders of $q$ and the WKB approach. Our findings demonstrate a convergence behavior, suggesting that the deviations are not caused by the order of $q$ or the WKB approach (and the currently used orders are sufficiently high to ensure the required accuracy of our conclusions).
Moreover, numerical analysis indicates that the errors arising from different choices of the orders of $q$ and the WKB approach are significantly smaller than the observed deviations. This provides evidence that the deviations are caused by the inherent differences in odd and even parities, rather than the order of $q$ or the WKB approach.

The results unequivocally demonstrate that, for a fixed nonlinear electromagnetic charge $q$, the axial and polar gravitational perturbations exhibit distinct QNM frequencies. This observation indicates a violation of isospectrality in the case of Bardeen de Sitter BHs. Unlike linear electromagnetic fields, the presence of nonlinear electromagnetic fields disrupts the isospectrality of gravitational perturbations.

Furthermore, we also compute the QNMs of the Bardeen Anti-de Sitter spacetime. To gain a clearer understanding of the impact of varying $q$ on the QNMs, we consider different ranges for the parameter $r_h$. Our results show the influence of the parameter $q$ on the real and imaginary parts of the QNMs.

It is worth noting that the master equations derived in this paper contain source terms. In this paper, due to calculation the QNMs, we impose the source terms vanish. However, if we consider the extreme mass ratio inspiral model, i.e., a point particle moving around the Bardeen BH, using our master equations one can calculate the energy flux at infinity or the gravitational wave of inspiral phase. On the other hand, the Bardeen BH considered in this paper is a simple RBH solution. The ABG solution seems to be a more reasonable choice as it approaches the Reissner-Nordstr{\"{o}}m solution at asymptotic infinity \cite{EAAG1}. Whether the breaking of isospectrality occurs in all nonlinear electromagnetic theories is worth further exploration. With the improvement of detection sensitivity, in the near future, LIGO/Virgo/KAGRA cooperation is expected to accurately measure the QNMs of the BH ringdown phase and confirm/deny the breaking of isospectrality from the QNM signal.

\section*{Acknowledgement}
Y. Zhao and W. Liu contributed equally to this work. This work was partially supported by the the Hunan Provincial Natural Science Foundation of China under Grant No. 2022JJ40262 and the National Natural Science Foundation of China under Grants No. 12375046 and No. 12205254.
~\\
\appendix
\section{Explicit Expressions for $E_{\mathrm{A}}$-$E_{\mathrm{K}}$}\label{appendixA}

By projecting the perturbed metric $h^{lm}_{ab}$ onto these orthogonal tensor bases, the expressions for coefficients A-K can be obtained.
\begin{eqnarray}
\mathrm{A}&=&f^2\oint h^{lm}_{ab}(v^av^bY^{\ast}_{lm})d\Omega,\non
\mathrm{B}&=&-\frac{f}{l(l+1)}\oint h^{lm}_{ab}v^{a}Y^{b\ast}_{E,lm}d\Omega, \non
\mathrm{C}&=&-\frac{f}{l(l+1)}\oint h^{lm}_{ab}v^{a}Y^{b\ast}_{B,lm}d\Omega,\non
\mathrm{D}&=&-\oint h^{lm}_{ab}v^{a}Y^{b\ast}_{R,lm}d\Omega,  \non
\mathrm{E}&=&\frac{1}{2}\oint h_{ab}^{lm}T^{ab\ast}_{T0,lm}d\Omega,\non\mathrm{F}&=&\frac{2(l-2)!}{(l+2)!}\oint h_{ab}^{lm}T^{ab\ast}_{E2,lm}d\Omega, \non
\mathrm{G}&=&\frac{2(l-2)!}{(l+2)!}\oint h_{ab}^{lm}T^{ab\ast}_{B2,lm}d\Omega,\non
\mathrm{H}&=&\frac{1}{l(l+1)f}\oint h^{lm}_{ab}T^{ab\ast}_{E1,lm}d\Omega, \non
\mathrm{J}&=&\frac{1}{l(l+1)f}\oint h^{lm}_{ab}T^{ab\ast}_{B1,lm}d\Omega,\non \mathrm{K}&=&\frac{1}{f^2}\oint h^{lm}_{ab}T^{ab\ast}_{L0,lm}d\Omega.
\end{eqnarray}
Only the case of $l\geq 2$ is considered in this paper.

Fild equation of ${E}_{A}-{E}_{K}$
\begin{eqnarray}
E_\mathrm{A}=&-&\frac{2(f-1+r^{2}\Lambda+2r^2\mathcal{L}+rf')}{r^2}\mathrm{A}\non
&-&\frac{(l^2+l+2f+4rf')f^2}{r^2}
\mathrm{K}-\frac{2f^3}{r}\frac{\pp }{\pp r}\mathrm{K}\non
&-&\frac{(l^2+l-2)f}{r^2}\mathrm{E}+\frac{f(6f+rf')}{r}\frac{\pp }{\pp r}\mathrm{E}
+2f^2\frac{\pp^2}{\pp r^2}\mathrm{E}~~~~
\end{eqnarray}
\begin{eqnarray}
E_\mathrm{B}=-\frac{f'}{r}\mathrm{D}-\frac{f}{r}\frac{\pp }{\pp r}\mathrm{D}-\frac{1}{r}
\frac{\pp }{\pp t}\mathrm{E}-\frac{f}{r}\frac{\pp }{\pp t}\mathrm{K},~~~~~~~~~~~~~~~~~~~~~~~~
\end{eqnarray}
\begin{eqnarray}
E_\mathrm{C}=&-&\frac{l^2+l-2+2r^{2}\Lambda+2f+2rf'+r^2f''+4r^2\mathcal{L}}{r^2}\mathrm{C}\non
&+&\frac{2f}{r}\frac{\pp}{\pp r}\mathrm{C}+f\frac{\pp^2}{\pp r^2}\mathrm{C}+\frac{3f}{r}\frac{\pp}{\pp t}\mathrm{J}+f\frac{\pp^2 }{\pp t\pp r}\mathrm{J},~~~~~~
\end{eqnarray}
\begin{eqnarray}
E_\mathrm{D}=&-&\frac{l^2+l-2+2r^{2}\Lambda+2f+4r^2\mathcal{L}+2rf'}{r^2}\mathrm{D}\non&+&\frac{rf'-2f}{rf}\frac{\pp}{\pp t}\mathrm{E}-2\frac{\pp^2}{\pp t\pp r}\mathrm{E}+\frac{2f}{r}\frac{\pp }{\pp t}\mathrm{K},~~~~~~~~~~~~~~~~~~~
\end{eqnarray}
\begin{eqnarray}
E_\mathrm{E}=&-&\frac{f(l^2+l+2rf'+2r^2f'')-r^2{f'}^2}{2r^2f^2}\mathrm{A}-\frac{rf'-2f}{2rf}\frac{\pp}{\pp r}\mathrm{A}\non
&+&\frac{\pp^2}{\pp r^2}\mathrm{A}+\frac{2f+rf'}{rf}\frac{\pp }{\pp t}\mathrm{D}+2\frac{\pp^2}{\pp t\pp r}\mathrm{D}\non
&-&\frac{4r\mathcal{L}+2f'+r(2\Lambda+2r\mathcal{L}'+f'')}{r}\mathrm{E}\non
&+&\frac{r^3\mathcal{L}'-2f-rf'}{r}\frac{\pp}{\pp r}\mathrm{E}-f\frac{\pp^2}{\pp r^2}\mathrm{E}+\frac{1}{f}\frac{\pp^2}{\pp t^2}\mathrm{E}\non
&+&\frac{r^2{f'}^2+f(l^2+l+6rf'+2r^2f'')}{2r^2}\mathrm{K}\non
&+&\frac{f(2f+rf')}{2r}\frac{\pp}{\pp r}\mathrm{K}+\frac{\pp^2}{\pp t^2}\mathrm{K},
\end{eqnarray}
\begin{eqnarray}
E_\mathrm{F}&=&-\frac{1}{r^2f}\mathrm{A}+\frac{f}{r^2}\mathrm{K},~~~~~~~~~~~~~~~~~~~~~~~~~~~~~~~~~~~~~~~~~~~~~~~~~~~~~~~~~~~~
\end{eqnarray}
\begin{eqnarray}
E_\mathrm{G}=-\frac{2}{rf}\frac{\pp }{\pp t}\mathrm{C}-\frac{2f+2rf'}{r^2}\mathrm{J}
-\frac{2f}{r}\frac{\pp}{\pp r}\mathrm{J},~~~~~~~~~~~~~~~~~~~~~~~~~~~~~~~
\end{eqnarray}
\begin{align}
E_\mathrm{H}&=\frac{2f+rf'}{2r^2f^2}\mathrm{A}-\frac{1}{rf}\frac{\pp}{\pp r}\mathrm{A}
-\frac{1}{rf}\frac{\pp}{\pp t}\mathrm{D}
+\frac{1}{r}\frac{\pp}{\pp r}\mathrm{E}
\non
&~~~~~-\frac{2f+rf'}{2r^2}\mathrm{K},~~~~~~~~~~~~~~~~~~~~~~~~~~~~~~~~~~~~~~~~~~~~~~~~~~~~~~~~~~~~~~~
\end{align}
\begin{align}
E_\mathrm{J}&=\frac{1}{rf}\frac{\pp}{\pp t}\mathrm{C}-\frac{1}{f}\frac{\pp^2}{\pp t\pp r}\mathrm{C}-\frac{1}{f}\frac{\pp^2}{\pp t^2}\mathrm{J}\non
&~~~~~-\frac{l^2+l-2+2r^{2}\Lambda+4r^2\mathcal{L}+2rf'+r^2f''}{r^2}\mathrm{J},~~~~
\end{align}
\begin{eqnarray}
E_\mathrm{K}=&-&\frac{l^2+l+2rf'}{r^2f^2}\mathrm{A}+\frac{2}{rf}\frac{\pp}{\pp r}\mathrm{A}
+\frac{4}{rf}\frac{\pp}{\pp t}\mathrm{D}+\frac{l^{2}+l-2}{r^{2}f}E\non
&-&\frac{2f+rf'}{rf}\frac{\pp}{\pp r}\mathrm{E}
+\frac{2}{f^2}\frac{\pp^2}{\pp t^2}\mathrm{E}\non
&-&\frac{2(2r^2\mathcal{L}-1+r^{2}\Lambda)}{r^2}
\mathrm{K},~~~~~~~~~~~~~~~~~~~~~~~~~~~~~~~~~~~~~~~~~~~~~~~
\end{eqnarray}

\section{Expressions For Parameters in \eqs{polareq} and \meq{polarsource}}\label{appendixB}

Here we present the explicit expressions of the parameters in \eq{polareq}, which are given by
\begin{align}
N=& -4f^{2}\mu-\sigma\left(-\rho\sigma+2f'\right)+2f\left(\gamma\sigma-\nu\sigma+\sigma'\right), \\
\sigma=& -2+\lambda +2r^{2}\Lambda +2f+2rf'+4r^{2}\kappa\mathcal{L}, \\
\tau=& -2+\lambda +2r^{2}\Lambda+3rf'+4r^{2}\kappa\mathcal{L}, \\
c_{0}=& \frac{1}{2fN\sigma^{2}}\bigg\{-2f^{2}M_{1}N\mu-M_{1}N\sigma f' \non
& +f\Big[\sigma N M_{1}'-\sigma M_{1}N'+M_{2} N\left(-\nu\sigma+\sigma'\right)\Big]\bigg\}, \\
c_{1}=& \frac{1}{2N \sigma}\bigg\{M_{1}N-f\Big[N^{2}+2f\sigma N' \non
& +2N\left(2f^{2}\mu+f\nu\sigma-\sigma f'-2f\sigma'\right)\Big]\bigg\},
\end{align}
and the parameters appearing in the source term \eq{polarsource} are given by

\begin{align}
M_{F}=& \frac{r}{\tau}\left(\eta+\lambda\sigma-2f\sigma+rf'\sigma\right), \\
N_{A}=& \frac{1}{4f^{2}N\sigma^{2}}\bigg\{rN^{2}+2rf\sigma N' \non
& +2N \Big[2rf^{2}\mu+2r\sigma f'+f\left(-\sigma+r\nu\sigma-r\sigma'\right)\Big]\bigg\}, \\
N_{F}=& \frac{1}{4fN\sigma^{2}\tau}\bigg\{rN^{2}\eta+2fM_{F}\sigma\tau N'+2N\tau\Big[2f^{2} M_{F}\mu  \non
& +M_{F}\sigma f'-f\left(\sigma M_{F}'-\nu\sigma M_{F}+\sigma' M_{F}\right)\Big]\bigg\},
\end{align}
Here we have,
\begin{align}
M_{1}=& f\Big[\rho\sigma^{2}+f\left(-2\nu\sigma+2\sigma'\right)\Big], \\
M_{2}=& 2f\left(2f^{2}\mu-f\gamma\sigma+\sigma f'\right), \\
\eta=& 4-\lambda ^{2}-8r^2\Lambda+4r^{4}\Lambda^{2}+4f^{2}+16r^{4}\kappa^{2}\mathcal{L}^{2}\non
& -8rf'-rf'\lambda+8r^{3}f'\Lambda+4r^{2}f'^{2} \non
& +16r^{2}\kappa\mathcal{L}\left(r^{2}\Lambda+rf'-1\right) \non
& +2f\left(-4+\lambda+4r^{2}\Lambda+8r^{2}\kappa\mathcal{L}+4rf'\right), \\
\gamma=& -\frac{1}{2f\left(\lambda-2+2r^{2}\Lambda+4r^{2}\kappa\mathcal{L}+3rf'\right)} \non
& \bigg\{4f\left(r\Lambda+2r\kappa\mathcal{L}+f'\right) \non
&  +f'\left(\lambda-2+2r^{2}\Lambda+4r^{2}\kappa\mathcal{L}+4rf'\right)\bigg\}, \\
\rho=& \frac{1}{r}-\frac{f'}{\lambda-2+2r^{2}\Lambda+4r^{2}\kappa\mathcal{L}+3rf'},
\end{align}
\begin{align}
\nu=& -\frac{1}{2rf\left(\lambda-2+2r^{2}\Lambda+4r^{2}\kappa\mathcal{L}+3rf'\right)} \times \non
& \bigg\{2f\left(\lambda-2+2r^{2}\Lambda+4r^{2}\kappa \mathcal{L}+4rf'\right)+ \non
&  rf'\Big[-rf'+5\left(\lambda-2+2r^{2}\Lambda+4r^{2}\kappa\mathcal{L}+3rf'\right)\Big]\bigg\} ,
\end{align}
\begin{align}
\mu=& \frac{1}{4rf^{2}\left(\lambda-2+2r^{2}\Lambda+4r^{2}\kappa\mathcal{L}+3rf'\right)} \times \non
& \bigg\{8f^{2}\left(\lambda-2+r^{2}\Lambda+2r^{2}\kappa\mathcal{L}+2rf'\right) \non
& -r^{2}f'^{2}\left(\lambda-2+2r^{2}\Lambda+4r^{2}\kappa\mathcal{L}+2rf'\right) \non
& +2f\Big[r^{2}f'^{2}+rf'\left(-12+6\lambda+2r^{2}\Lambda-3r^{2}f''\right)   \non
& +\left(\lambda-2+2r^{2}\Lambda\right)\left(2\lambda-4-r^{2}f''\right)  \non
& -4r^{2}\kappa\mathcal{L}\left(4-2\lambda-rf'+r^{2}f''\right)\Big]\bigg\} .
\end{align}

\begin{widetext}
\section{Quasinormal Modes Frequencies}\label{appendixC}

\begin{table}[h]

\renewcommand{\arraystretch}{1.2}
\centering
\caption{QNM frequencies of the gravitational perturbations calculated by WKB approach with different $l$ and $n$ in Badeen de Sitter spacetime, the parameters are chosen as $q=0.1$ and $\Lambda=0.02$.}
\label{tab:1}
\setlength\tabcolsep{7.2mm}{
\begin{tabular}{cccccccc}
\hline\hline
\multirow{2}{*}{$l$}	&
\multirow{2}{*}{$n$}  &
\multicolumn{2}{c}{odd parity} & \multicolumn{2}{c}{even parity}& \multicolumn{2}{c}{relative deviation} \\
\cline{3-4}  \cline{5-6}  \cline{7-8}
&~&{\bf Re($\omega$)}&{\bf -Im($\omega$)}&{~\bf Re($\omega$)}&{\bf -Im($\omega$)}&{~\bf Re($\omega$)}&{\bf -Im($\omega$)}\\
\hline
\multirow{2}{*}{2}&0
&0.3396  &0.0816
&0.3392  &0.0817
&0.1178\%  &0.1225\%
\\
&1 &0.3201  &0.2485
&0.3196    &0.2490
&0.1562\%   &0.2012\%
\\
\hline
\multirow{3}{*}{3}&0
&0.5446   &0.0844
&0.5443  &0.0844
&0.0551\%  &0\%
\\
&1 &0.5323  &0.2551
&0.5320   &0.2552
&0.0564\%  &0.0392\%
\\
&2 &0.5087  &0.4316
&0.5083   &0.4316
&0.0786\%   &0\%
\\
\hline
\multirow{4}{*}{4}&0
&0.7348  &0.0855
&0.7346  &0.0855
&0.0272\%  &0\%
\\
&1  &0.7256  &0.2577
&0.7253   &0.2577
&0.0413\%  &0\%
\\
&2  &0.7076  &0.4331
&0.7073   &0.4331
&0.0424\%  &0 \%
\\
&3  &0.6818 &0.6142
&0.6816  &0.6142
&0.0293\%  &0 \%
\\
\hline
\multirow{5}{*}{5}&0
&0.9190   &0.0861
&0.9188   &0.0861
&0.0218\%  &0\%
\\
&1  &0.9116   &0.2589
&0.9114   &0.2589
&0.0219\%  &0\%
\\
&2  &0.8970  &0.4339
&0.8968   &0.4339
&0.0223\%  &0 \%
\\
&3  &0.8758  &0.6126
&0.8756  &0.6126
&0.0228\%  &0 \%
\\
&4  &0.8488 &0.7964
&0.8485   &0.7964
&0.0353\% &0 \%
\\
\hline\hline
\end{tabular}}
\end{table}
	
\begin{table}[h]
\renewcommand{\arraystretch}{1.2}
\centering
\caption{QNM frequencies of the gravitational perturbations calculated by WKB approach with different $q$ and $\Lambda$ in Badeen de Sitter spacetime, here $l=2$ and $n=0$.}
\label{tab:2}
\setlength\tabcolsep{6.8mm}{
\begin{tabular}{cccccccc}
\hline\hline
\multirow{2}{*}{$q$}	& \multirow{2}{*}{$\Lambda$}  &
\multicolumn{2}{c}{odd parity}&
\multicolumn{2}{c}{even parity}&
\multicolumn{2}{c}{relative deviation} \\
\cline{3-4} \cline{5-6} \cline{7-8}
&~&{\bf Re($\omega$)}&{\bf -Im($\omega$)}&{\bf Re($\omega$)}&{\bf -Im($\omega$)}&{\bf Re($\omega$)}&{\bf -Im($\omega$)}\\
\hline
\multirow{6}{*}{0.2}&0.00
&0.3784    &0.0883
&0.3766    &0.0886
&0.4757\%  &0.3238\%
\\
&0.02&0.3432 &0.0813
&0.3415    &0.0815
&0.4953\%  &0.2472\%
\\
&0.04&0.3039 &0.0731
&0.3023    &0.0732
&0.5265\%  &0.1752\%
\\
&0.06&0.2586 &0.0631
&0.2572    &0.0632
&0.5414\%  &0.1220\%
\\
&0.08&0.2035 &0.0504
&0.2024    &0.0505
&0.5405\%  &0.1646\%
\\
&0.10&0.1265 &0.0318
&0.1258    &0.0318
&0.5534\%  &0\%
\\
\hline
\multirow{6}{*}{0.4}&0.00 &0.3935
&0.0868 &0.3859 &0.0873
&1.9314\% &0.5760\%
\\
&0.02 &0.3588 &0.0801 &0.3515 &0.0800
&2.0346\% &0.1248\%
\\
&0.04 &0.3202 &0.0724 &0.3134 &0.0728
&2.1237\% &0.5525\%
\\
&0.06&0.2760 &0.0633 &0.2699 &0.0636
&2.2101\% &0.4739\%
\\
&0.08&0.2230 &0.0519 &0.2179 &0.0520
&2.2870\% &0.1927\%
\\
&0.10&0.1527 &0.0360 &0.1491 &0.0361
&2.3576\%&0.2778\%
\\
\hline
\multirow{6}{*}{0.6}&0.00&0.4282
&0.0815 &0.4049 &0.0839
&5.4414\% &2.9448\%
\\
&0.02&0.3927 &0.0765 &0.3718 &0.0780
&5.3221\% &1.9608\%
\\
&0.04&0.3542 &0.0703 &0.3355 &0.0711
&5.2795\% &1.1380\%
\\
&0.06&0.3113 &0.0628 &0.2947 &0.0632
&5.3325\% &0.6369\%
\\
&0.08&0.2616 &0.0535 &0.2475 &0.0537
&5.3899\% &0.3738\%
\\
&0.10&0.1999 &0.0414 &0.1890 &0.0414 &5.4527\% &0\%
\\				
\hline\hline
\end{tabular}}
\end{table}
	
\begin{table}[h]
\renewcommand{\arraystretch}{1.2}
\centering
\caption{The axial gravitational QNM frequencies of the Bardeen Anti-de Sitter spacetime calculated by the HH method with different $q$ and $r_h$, where the cosmological constant is set as $\Lambda=-3$, and the cut off number of the summation is determined as $N=50$.}
\label{tab:4}
\setlength\tabcolsep{4.5mm}{
\begin{tabular}{cccccccccc}
\hline\hline
\multirow{2}{*}{$q$}&
\multirow{2}{*}{\textbf {$r_{h}$}}&
\multicolumn{2}{c}{$l=2, n=0$}&
\multicolumn{2}{c}{$l=2, n=1$}&
\multicolumn{2}{c}{$l=3,n=0$}&
\multicolumn{2}{c}{$l=3,n=1$} \\
\cline{3-4}\cline{5-6}\cline{7-8}\cline{9-10}
&~&{\bf Re($\omega$)}&{\bf -Im($\omega$)} &{\bf Re($\omega$)}&{\bf -Im($\omega$)}&{\bf Re($\omega$)}&{\bf -Im($\omega$)}&{\bf Re($\omega$)}&{\bf -Im($\omega$)} \\
\hline
\multirow{7}{*}{0.2}&2&0\texttildelow&0.816839&4.36465&5.38995
&0\texttildelow&2.38847&4.47733&5.22429
\\
&4&0\texttildelow&0.369257&7.76816&10.6822
&0\texttildelow&0.913702&7.93976&10.6235
\\
&6&0\texttildelow&0.242054&11.3485&16.0005
&0\texttildelow&0.587737&11.4762&15.9616
\\
&8&0\texttildelow&0.180492&14.9857&21.3237
&0\texttildelow&0.435592&15.0850 &21.2945
\\
&10&0\texttildelow&0.144009&18.6470&26.6488
&0\texttildelow&0.346595&18.7277 &26.6254
\\
&50&0\texttildelow&0.028672&92.5018&133.195
&0\texttildelow&0.068692&92.5184&133.190
\\
&100&0\texttildelow&0.014334&184.957&266.386
&0\texttildelow&0.034337&184.966 &266.384
\\
\hline
\multirow{7}{*}{0.4}&2&0\texttildelow&1.10090
&4.10727 &5.91645
&0\texttildelow&3.29842
&3.88143 &6.03826
\\
&4&0\texttildelow&0.455923
&7.65404 &10.8665
&0\texttildelow&1.01260
&7.82060 &10.8118
\\
&6&0\texttildelow&0.295336
&11.2736 &16.1174
&0\texttildelow&0.643960
&11.4004 &16.0793
\\
&8&0\texttildelow&0.219352
&14.9298 &21.4100
&0\texttildelow&0.475610
&15.0288 &21.3810
\\
&10&0\texttildelow&0.174700
&18.6023 &26.7173
&0\texttildelow&0.377861
&18.6829 &26.6940
\\
&50&0\texttildelow&0.034677
&92.4929 &133.208
&0\texttildelow&0.074702
&92.5095 &133.204
\\
&100&0\texttildelow&0.017335
&184.953 &266.393
&0\texttildelow&0.037338
&184.961 &266.391
\\
\hline
\multirow{7}{*}{0.6}&2&0\texttildelow&1.64729
&4.34131 &7.41637
&0\texttildelow&\texttildelow&5.19887 &7.46867
\\
&4&0\texttildelow&0.601735
&7.45796 &11.2095
&0\texttildelow&1.18029
&7.61522 &11.1654
\\
&6&0\texttildelow&0.384439
&11.1468 &16.3213
&0\texttildelow&0.738091
&11.2720 &16.2847
\\
&8&0\texttildelow&0.284234
&14.8357 &21.5573
&0\texttildelow&0.542447
&14.9342 &21.5289
\\
&10&0\texttildelow&0.225908
&18.5275 &26.8332
&0\texttildelow &0.430034
&18.6079 &26.8102
\\
&50&0\texttildelow&0.044687
&92.4781 &133.231
&0\texttildelow&0.084719
&92.4947 &133.226
\\
&100&0\texttildelow&0.022336
&184.946 &266.404
&0\texttildelow&0.042340
&184.954 &266.402
\\
\hline\hline
\end{tabular}}
\end{table}
	
\begin{table}[t]
\renewcommand{\arraystretch}{1.2}
\centering
\caption{The polar gravitational QNM frequencies of the Bardeen Anti-de Sitter spacetime calculated by the HH method with different $q$ and $r_h$, where the cosmological constant is set as $\Lambda=-3$, and the cut off number of the summation is determined as $N=150$.}
\label{tab:5}
\setlength\tabcolsep{10.5mm}{
\begin{tabular}{cccccc}
\hline\hline
\multirow{2}{*}{$q$}&
\multirow{2}{*}{\textbf {$r_{h}$}}&
\multicolumn{2}{c}{$l=2,~n=0$}&
\multicolumn{2}{c}{$l=3,~n=0$} \\
\cline{3-4}\cline{5-6}
&~&{\bf Re($\omega$)}&{\bf -Im($\omega$)} &{\bf Re($\omega$)}&{\bf -Im($\omega$)} \\
\hline
\multirow{7}{*}{0.2}&2&4.48199
&3.95046&4.58341&3.30879
\\
&4&8.03115&9.55387&8.32398&8.57327
\\
&6&11.5906&14.9412&11.9309&14.3709
\\
&8&15.2177&20.2144&15.4865&19.9298
\\
&10&18.8806&25.4391&19.0817&25.3612
\\
&50&93.1610&128.669&92.8456&130.662
\\
&100&186.243&257.431&185.524&261.567
\\	
\hline				
\multirow{7}{*}{0.4}&2&4.39016
&3.82357 &4.53669 &3.32935
\\
&4&8.20505 &8.94905 &8.20260 &8.38047
\\
&6&12.0580 &13.9709 &11.8863 &13.8573
\\
&8&15.9442 &18.9028 &15.5437 &19.1484
\\
&10&19.8492 &23.7920 &19.2304 &24.3338
\\
&50&98.5247 &120.387 &94.3123 &125.134
\\
&100&197.003 &240.864 &188.5045 &250.488
\\		
\hline\hline
\end{tabular}}
\end{table}
\end{widetext}

\end{document}